\documentclass[manuscript]{aastex}
\usepackage{graphicx}

\begin{document}

\title{Dynamics of the giant planets of the solar system in the
  gaseous proto-planetary disk and relationship to the current
  orbital architecture}

\author{\textbf{\large Alessandro  Morbidelli}}
\affil{\small\em Observatoire de la C\^ote d'Azur, Nice, France}

\author {\textbf{\large Kleomenis Tsiganis}}
\affil{\small\em Dep.
of Physics, Aristotle University of Thessaloniki, 
Greece}

\author {\textbf{\large Aur\'elien Crida}}
\affil{\small\em Dep. of Physics, Univ. of Tubingen, Germany}

\author {\textbf{\large Harold F. Levison}}
\affil{\small\em Southwest Research Institute, Boulder, Colorado}

\author {\textbf{\large Rodney Gomes}}
\affil{\small\em National Observatory of Rio de Janeiro, Bresil}

\author{\hbox{}}

\author{Corresponding author:\\
Alessandro Morbidelli\\
OCA\\
B.P. 4229\\
06304 Nice Cedex 4, France\\
email: morby@obs-nice.fr}

\received{June 7, 2007}
\accepted{July 17, 2007}

%\newpage

\begin{abstract}

{We study the orbital evolution of the 4 giant planets of our
  solar system in a gas disk. Our investigation extends the previous
  works by Masset and Snellgrove (2001) and Morbidelli and Crida
  (2007, MC07), which focussed on the dynamics of the Jupiter-Saturn
  system. The only systems that we found to reach a steady state are
  those in which the planets are locked in a quadruple mean motion
  resonance (i.e. each planet is in resonance with its neighbor). In
  total we found 6 such configurations.  For the gas disk parameters
  found in MC07, these configurations are characterized by a
  negligible migration rate. After the disappearance of the gas, and
  in absence of planetesimals, only two of these six configurations
  (the least compact ones) are stable for a time of hundreds of
  millions of years or more. The others become unstable on a timescale
  of a few My. Our preliminary simulations show that, when a
  planetesimal disk is added beyond the orbit of the outermost planet,
  the planets can evolve from the most stable of these configurations
  to their current orbits in a fashion qualitatively similar to that
  described in Tsiganis et al. (2005).}

\end{abstract}


\section{Introduction}

Of all the planetary systems known to date, our solar system remains
undoubtedly the one for which we have the largest number of accurate
observational constraints {to be used for modeling the} evolution of
the giant planets {up to their current orbital
configuration}. Reconstructing this evolution as far back in time as
possible is very important, because it can constrain the orbits on
which the planets formed, and in turn, shed new light onto their
formation mechanism. In particular, { knowledge of} these aspects may
allow us to understand why our solar system looks so different from
the extra-solar systems discovered so far. Two differences are
particularly striking. { First, our} giant planets are all far from
the Sun, {whereas giant planets in the very vicinity of their host
stars are numerous in extra-solar systems. Observational biases favor
the discovery of these planets, but the very fact that they exist in
other systems and not in our own is real and remarkable.} Thus, we
need to understand in which conditions planets can avoid large-range
radial migration towards the central star. 
{ Second,} the orbital eccentricities of
extra-solar planets, including those at distances of several AUs from
the central star, are { generally} much larger than the eccentricities
of the giant planets of our system. The latter, equal to several
percent, are nevertheless non-negligible. Planet eccentricities are
believed to be the result of mutual perturbations (Rasio and Ford, 1996;
Marzari and Weidenschilling, 2002). It is important to understand which
perturbations are responsible for the moderate eccentricities of the
giant planets of the Solar System and why the orbital excitation could be
much stronger in most extra-solar cases. 

Our group has recently proposed two models that aim to reconstruct two
different phases of the
evolution of the  
solar system: the one that was dominated by the gas disk
and the one that occurred after the disappearance of the gas. 

The first model addresses specifically the migration of Jupiter and Saturn
  in the { proto-planetary} gas disk. {If considered individually},
  these planets should have { evolved towards the Sun, as the result
  of Type~II migration}. However, Masset and Snellgrove (2001, MS01
  hereafter) showed that Saturn tends to get locked in the 2:3 MMR
  with Jupiter. In this configuration, the gaps opened { in the disk}
  by the two planets can overlap with each other. This can lead to a
  reversal of the migration direction.  This mechanism has more
  recently been studied in greater details by Morbidelli and Crida
  (2007, MC07 hereafter). They showed that, in the gas-disk parameter
  space represented by viscosity and scale height, there is a
  one-parameter family of solutions such that, once locked in the 2:3
  MMR, Jupiter and Saturn do not migrate. Sets of parameters close to
  this family lead to inward or outward migration, but with rates that
  are much slower than the { theoretically predicted} Type~II
  migration rates {for a single planet}. MC07 also showed that this
  kind of non-migrating, or slowly migrating, evolution is possible
  only if the planets involved have a mass ratio close to that of
  Jupiter and Saturn. Planets of similar masses, or - worse - systems
  where the outer planet is the more massive, inevitably lead to a
  fast inward migration. Therefore, they argued that the absence of a
  hot/warm Jupiter in our solar system is due to the specific mass
  hierarchy of our giant planets and to their formation on {
  initially} close-by orbits. To support this claim they pointed out
  that none of the known extra-solar planetary systems with two bodies
  close to their parent stars fulfill the conditions { necessary to}
  avoid Type~II migration: either the planets have comparable masses,
  or the outermost one is the most massive, or they are too separated
  to have { sculpted} overlapping gaps in their primordial gas disks.

The second model (Tsiganis et al., 2005; Gomes et al., 2005), which
actually was developed first, is often called the `Nice model' because
it was developed by an international collaboration at the Nice
Observatory in France. This model argued that, {if the giant planets
had a more compact configuration at the end of the gas-disk phase,
their subsequent migration driven by the interaction with a
planetesimals disk could have forced them to cross some mutual mean
motion resonance (MMR), thereby triggering a global instability; the
current orbital configuration could then be achieved from the
gravitational interaction between the planets and the disk particles.}
More precisely, the Nice model postulated that {the ratio of the
orbital periods of Saturn and Jupiter was initially slightly less than
2, so that the planets were close to their} mutual 1:2 MMR; Uranus and
Neptune { were supposedly orbiting the Sun a few AUs beyond the gas
giants}, and a massive planetesimal disk was extending from about
1.5~AU beyond the last planet up to 30--35~AU.  As a consequence of
the interaction of the planets with the planetesimal disk, the giant
planets suffered orbital migration, which { slowly} increased their
orbital separation. { As shown in their N-body simulations,} after a
long { quiescent phase} (with a duration varying from 300~My to 1~Gy,
depending on the exact initial conditions), Jupiter and Saturn { were
forced to} cross their mutual 1:2 MMR. This event excited their
orbital eccentricities to values similar to those presently
observed. The acquisition of eccentricity by { both} gas giants
destabilized Uranus and Neptune. Their orbits became very eccentric,
so that they penetrated deep into the planetesimal disk. Thus, the
planetesimal disk was dispersed, and the interaction between planets
and planetesimals finally parked { all four} planets on orbits with
separations, eccentricities and inclinations similar to what we
currently observe. This model has a long list of successes. As {
already} said, it explains the current orbital architecture of the
giant planets (Tsiganis et al., 2005).  It also explains the origin of
the so-called Late Heavy Bombardment (LHB), {a spike in the cratering
history of the terrestrial planets that occurred $\sim 650$~My after
planet formation. In the Nice model, the LHB is triggered by the
dispersion of the planetesimal disk; both the timing, the duration and
the intensity of the LHB deduced from Lunar constraints are well
reproduced by the model} (Gomes et al., 2005). { Furthermore,} {the
Nice model} also explains the capture of planetesimals around the
Lagrangian points of Jupiter, with a total mass and orbital
distribution consistent with the observed Jupiter Trojans (Morbidelli
et al., 2005). More recently, it has been shown to provide a framework
{ for understanding} the capture and orbital distribution of the
irregular satellites of Saturn, Uranus and Neptune (Nesvorny et al.,
2007). The main properties of the Kuiper belt (the relic of the
primitive trans-planetary planetesimal disk) have also been explained
in the context of the Nice model (Levison et al., 2007; see Morbidelli
et al., 2007, for a review).

The problem {we are now facing} is that a 2:3 MMR configuration of
Jupiter and Saturn, advocated to explain why the planets did not
migrate { towards} the Sun, is {different from} the initial conditions
of the Nice model, in which Jupiter and Saturn have initially
non-resonant orbits, { $\sim 0.5$~AU interior to} the 1:2 MMR. MC07
proposed a few mechanisms to extract Saturn from the 2:3 MMR with
Jupiter and bring it close the the 1:2 MMR at the end of the gas disk
phase.  However, we believe that a more complete system
(i.e.\ one with Uranus and Neptune) would probably become 
immediately unstable when this happens, in contrast
with the Nice model.

{This paper is the first in a series of two in which} we explore
this problem more thoroughly. In particular, we try to bridge the
results of MS01/MC07 with the fundamental aspects of the Nice
model. In section~\ref{4pl-gas} we start from a configuration found
in MC07 where Jupiter and Saturn are not migrating and we progressively add
Uranus and Neptune into the problem. {Following this procedure}, we
find 6 fully resonant relative configurations of the 4 planets, which
are stable and avoid migration towards the Sun. In
section~\ref{gas-disp}, we study the stability of these configurations
on longer timescales, after the gas disk {has disappeared
  and in absence of a planetesimals disk}.  We
find that 2 of the 6 configurations are stable over very long
time (several $10^8$ to $10^9$ years).
% and therefore {can be} 
%compatible with the onset of a late instability at the LHB time,
%driven by the interaction { of the planets} with a trans-planetary
%planetesimal disk. The other { 4 configurations} may lead Jupiter { to
%develop} an eccentric orbit, quite similar to those of the extra-solar
%planets { discovered so far at} the largest distances from their
%parent stars. 
{Although a detailed analysis of the evolution of these configurations
  under the influence of a planetesimal disk is left to the
  fore-coming paper, as a proof of concept, we present a couple of
  simulations in section~\ref{Nice2},} showing that the planets could
  eventually achieve an orbital architecture similar to their current
  one. The proposed evolution is different in the details from that of
  the Nice model (different resonances are involved in triggering the
  planetary instability) but the { basic mechanism and evolution are
  the same. The conclusions of this paper follow, in the last section.

\section{Four planet dynamics in the gas disk}
\label{4pl-gas}

We use the hydro-dynamical code developed by Crida et al.~(2007) on
the basis of the FARGO code by Masset (2000a,b), to simulate the
dynamics of the planets in the proto-planetary gas disk. In the Crida
et al. scheme, the disk is represented using a { combination} of 2-D
and 1-D grids. The main part of the disk, in which the planets evolve,
is { sampled by} a 2-D grid in polar barycentric coordinates.  This
grid extends from $r=0.3$ to 5 in radius (in units of the initial
orbital radius of Jupiter) and has a resolution of 282 in radius and
325 in azimuth. The planets also evolve on coplanar orbits.  The inner
part of the disk (ranging from $r=0.016$ to the innermost boundary of
the 2-D grid) and the outer part of the disk (ranging from the
outermost boundary of the 2-D grid to $r=40$) are { sampled by} a 1-D
grid.  These 1-D grids have open outflow boundaries at $r=0.016$ and
$r=40$, while they exchange information with the 2-D grid at their
common boundaries in order to supply realistic, time-dependent,
boundary conditions for the latter. The algorithm for this interface
between the 1-D and 2-D grids { explicitly requires} that the angular
momentum of the global system (the disk in the 2-D section, plus the
disk in the 1-D section plus the planets-star system) is
conserved. With this approach, the global viscous evolution of the
disk and the local planet-disk interactions are both described well
and the feedback of one on the other is properly taken into
account. Because the migration of the giant planets depends on the
global evolution of the disk, this code provides more realistic
results than the usual algorithms in which the evolution of the
considered { 2D} portion of the disk depends crucially on the adopted
(arbitrary) boundary conditions. {For more information and accuracy
tests we refer the reader to Crida et al. (2007).}

{We adopt a set of disk parameters} from MC07 in which Jupiter and
Saturn did not migrate after they became locked in the 2:3 MMR. The
scale height of this disk is 5\% and its viscosity (constant over
radius) is $\nu=3.2\times 10^{-6}$, assuming the Sun's mass and the
initial semi-major axis of Jupiter are the units of mass and distance,
respectively. In the usual $\alpha$ prescription (Shakura and Sunyaev,
1973), this viscosity corresponds to $\alpha=1.2\times 10^{-3}$ at
$r=1$. The initial surface density of the disk is $\Sigma(r)= 3\times
10^{-4} \exp(-r^2/53)$. This $\Sigma$ was {inspired by the results} of
Guillot and Hueso (2006) {who studied the structure of a disk that
viscously evolved under the effects of the collapse of fresh matter from
the proto-stellar cloud, of the gas  viscous spreading
and of photo-evaporation by the central and neighboring stars}. {The
exact choice of $\Sigma(r)$ should not be crucial for our analysis,
although some issues should be kept in mind.  $\Sigma$ is a
multiplicative factor in the equations of motion, so it simply governs
the evolution timescale. In case of differential migration of multiple
planets, the value of $\Sigma$ --here close to the minimal mass solar
nebula (Hayashi, 1981)-- determines the relative migration rates,
and hence the probability of capture in the various mutual resonances.
The radial profile of $\Sigma(r)$ also affects the relative migration
rates of planets at different locations. However, given that in our
simulations (see below) all the planets are within a factor of 2.5 in
heliocentric distance, the sensitivity of their relative evolution on
the radial profile of $\Sigma$ should be moderate. We will come back
to these issue when discussing our results.}

Fig.~\ref{MC07fig}, reproduced from MC07, shows the evolution of the
semi-major axes of Jupiter and Saturn, after they became locked in the
2:3 MMR, over 1,500 Jovian orbital periods. A slight parallel outward
migration is visible, which could be annealed with a slight increase
of the disk's scale height or viscosity.  The eccentricities have only
small-amplitude oscillations around a { small} constant value (0.015
for Saturn and 0.004 for Jupiter). We refer to MS01 and MC07 for { an
explanation} of why Type~II migration is prevented in this
configuration. We just stress { here} that this mechanism is robust.
For a given (reasonable) viscosity, the coupled Jupiter-Saturn pair
migrates outward in a thin disk, while it moves inward in a thick
disk. Thus, it is always possible to find a disk scale height for which
the migration vanishes. If some simulation parameters are changed (for
instance the prescription of the boundary conditions, the radial
dependence of the viscosity, or the scale over which the potential of
each planet is smoothed - here set to 0.7 $H$, where $H$ is the local height
of the disk at the planet's position - the exact value of the disk
scale height that allows for a non-migrating solution may change, but
the very existence of such a solution is not at risk.

The fact that Jupiter did not migrate closer to the Sun argues that
the {inward} migration of the Jupiter-Saturn pair was, for the
most part, inhibited. On the other hand, it is difficult to believe
that Jupiter and Saturn migrated outward because the asteroid belt
would have been completely decimated if Jupiter had been closer to the
Sun. (Note that models that assume that Jupiter was, more or less, at
its current location adequately reproduce the observed structure of
the asteroid belt; see Petit et al., 2002, for a review). Thus, the
philosophy of MC07, as well as of this work, is to construct models
where the structure the proto-planetary disk is such that the planets
do not migrate significantly.  Given that this happens in our numerical
scheme when $H/r=0.05$ and $\nu=3.2\times 10^{-6}$ (for instance), we
simply adopt these parameters with the understanding that that the
real disk may have been different.

We now extend the work of MC07 by adding Uranus and Neptune to the
calculation.  Given that the heliocentric order of the ice giants
changed during $\sim\!50\%$ of the successful simulations of the
original Nice model -- so that we don't know which one formed closer
to the Sun -- we assume that the two planets have the same mass: 15
Earth masses each. For simplicity, we nevertheless call the innermost
ice giant `Uranus,' and the outermost one `Neptune.'

The goal of this section is to find stable configurations for the four
planets.  To accomplish this we employ the following procedures. We
pick up the MC07 simulation shown in Fig.~\ref{MC07fig} at $t=1,300$,
at which point we introduce Uranus into the calculation.  In contrast
to the procedures used by MC07 for Jupiter and Saturn, we let Uranus
migrate freely from the moment that it is introduced into the
simulation.  We think that this is a legitimate change because
Uranus's effect on the disk's surface density profile is minimal and
occurs on a timescale that is short compared to Uranus's migration
timescale.

In the first simulation, Uranus was initially placed at $r=2.55$,
which is beyond the 1:2 resonance with Saturn.  Uranus moves inward
relatively rapidly due to Type~I migration.  It jumps across the 1:2
resonance with Saturn (the resonance is too weak to capture it given
the migration speed generated by our assumed disk; see also MS01) and
is eventually trapped in the 2:3 resonance with Saturn
(Fig.~\ref{Uranus}, black curve).  In the second simulation, we start
Uranus at $r=1.90$, which is between the 2:3 and 1:2 resonances with
Saturn. Again, the planet gets trapped in the 2:3 resonance
(Fig.~\ref{Uranus}, dark gray curve). In both simulations, the capture
into resonance increases the eccentricity of Uranus from $\sim 0$ to
$\sim 0.025$.  After this capture, the three planets evolve in
parallel, showing that they have reached a stable relative
configuration in a three-body 2:3+2:3 resonance. 

Notice that after this configuration is reached, Jupiter's and
Saturn's outward migration is accelerated somewhat.  We might expect
the opposite since Uranus feels a negative torque from the disk that
should be transmitted to Jupiter and Saturn through the resonances.
The slow outward migration is due to the fact that Uranus slightly
depletes the disk outside of Saturn's orbit. As a consequence, the
balance of the torques exerted on Jupiter and Saturn by the disk is
broken, and the positive torque felt by Jupiter dominates.  However,
as we said above, this does not invalidate the general MS01/MC07
scenario because we could restore the torque equilibrium if we were to
slightly increase in the scale height of the disk.

In a third simulation, we started Uranus at $r=1.73$, which is between
the 2:3 and 3:4 resonances with Saturn. Again, we observe an inward
drift due to Type~I migration until the planet is trapped in Saturn's
3:4 MMR.  After this, the relative configuration of the three planets
does not change (see Fig.~\ref{Uranus}, gray curve), although, as
described above, the whole system migrates outward.  The eccentricity
of Uranus does not exceed 0.01.  The evolution of Jupiter and Saturn
are indistinguishable in all three simulations, and so we just plot
those of the first simulation for clarity.

Furthermore, we perform a final simulation were we place Uranus
initially at $r=1.61$, which is between the 3:4 and 4:5 MMR with
Saturn. In this case, the evolution is different from those described
above. In particular, the motion of Uranus is unstable due to its
proximity to the gas giants.  Hence, Uranus is pushed outward until it
finds again a stable relative configuration in the 3:4 MMR with Saturn
(see Fig.~\ref{Uranus}, light gray curve).  Its subsequent evolution
is indistinguishable from the previous one.

From the above three experiments, we deduce that, for our assumed
disk, there are two stable and invariant configurations of the
three-planet system: Uranus being either in the 2:3 or the 3:4 MMR
with Saturn. 

Next, we introduce Neptune into the problem. We start by considering
the first of the aforementioned simulations, where Uranus was trapped
in the 2:3 MMR with Saturn. We continue this simulation after placing
Neptune at $r=2.58$, i$.$e$.$ between the 2:3 and 1:2 MMR with Uranus.
As expected, it drifts inward due to Type~I migration
(Fig.~\ref{U23Neptune}, black curve).  However, Neptune is not trapped
in Uranus's 2:3 MMR, but crosses it, because the resonance is too weak
to trap a body at Neptune's migration speed. Neptune is, however,
subsequently trapped in Uranus's 3:4 MMR.  Note that the 3:4 MMR with
Uranus is also the 1:2 MMR with Saturn since Uranus and Saturn are in
the 2:3 MMR. The capture of Neptune into resonance pumps the
eccentricity of Uranus up to about 0.05, whereas the eccentricity of
Neptune increases only to $\sim 0.01$.

Repeating the simulation with Neptune starting from $r=2.34$ (between
the 3:4 and 2:3 MMR with Uranus), also leads to capture in the 3:4 MMR
with Uranus.  After the trapping, the evolution of the two systems are
indistinguishable (Fig.~\ref{U23Neptune}, dark gray curve).
Conversely, starting the simulation with Neptune at $r=2.19$ or
$r=2.11$, leads to its capture into the 4:5 and 5:6 MMRs with Uranus,
respectively (gray and light gray curves in Fig.~\ref{U23Neptune}).
The eccentricities of the ice giants are progressively smaller with
increasing $m$, for a $m:m+1$ resonance.  For the 5:6 MMR, the
eccentricity of Uranus and Neptune become $\sim 0.04$ and $\sim
0.007$, respectively.

We repeat the same exercise, but this time in the system where Uranus
was trapped in the 3:4 MMR with Saturn, and find similar results (see
Fig.~\ref{U34Neptune}). The evolution of the eccentricities (not shown
in the figure) is also similar to what is observed in the runs of
Fig.~\ref{U34Neptune}.  Thus, we conclude that for each of the two
stable Jupiter-Saturn-Uranus configurations, there are three stable
locations for Neptune: in Uranus's 3:4, 4:5 or 5:6 MMR.  Thus, we
found 6 planetary configurations in total.  In all cases, the four
planets form a fully resonant system. In addition, all these
configurations are characterized by an almost complete absence of
radial migration, which is required to explain the absence of a
hot/warm Jupiter in our solar system.

It is important to keep in mind that we cannot be sure that the 6
configurations found here are the only possible final states for the
problem at hand.  For example, we might reach a different
configuration if we were to introduce Uranus and Neptune at the same
time and on mutually scattering orbits in the vicinity of Jupiter or
Saturn.  Similarly, if we were to use a lower mass disk, the ice giants
might be captured into weaker resonances -- e.g.\ the 1:2 resonance
with Saturn (for Uranus) or the 2:3 resonance with Uranus (for
Neptune).  However, these systems will be less compact than our six,
and, given the results in section~\ref{Nice2}, we believe that they
are unlikely to evolve into systems resembling the real giant planets.
Thus, for the remainder of this paper we will restrict our analysis to
the six configurations found in this section.

\section{Evolution of the planets after the disappearance of the gas disk}
\label{gas-disp}

In order to test the long-term stability of the planetary
configurations constructed in the last section, we first have to
transition from a gaseous to a gas-free environment. Gas disks are
typically dispersed on a timescale of $10^5$--$10^6$~y (Haisch et al.,
2001).  Unfortunately, we do not yet know how this dispersal
took place.  Photo-evaporation is probably the key, but exactly how it
proceeds (i$.$e$.$ from the inside first, as argued in Alexander et
al., 2006, or from the outside first, as found by Adams et al., 2004)
it is still debated.  Thus, we decided to implement a very simple
transition in our hydro-dynamical code, where we do not change the
shape of the disk's surface density profile, but decrease the total
amount of gas exponentially in time.  We chose a decay rate such that
the gas mass is halved in 160 ``Jovian'' orbital periods (i.e.\ at
$r=1$).  These hydro-dynamical are performed for 1,500 Jovian orbital
periods, implying that the disk is reduced by a factor 670 at the end
of the simulation, i$.$e$.$ there is effectively no gas left.

Note that we are not claiming that the disappearance of the gas disk
actually followed this simple recipe. Our aim is only to change the
disk potential as smoothly and slowly as possible (given the available
computing time) in order to give the planets enough time to adapt to
the evolving situation.  During these simulations we do not observe
any significant evolution in the semi-major axes of the planets, and
so the resonant structure is preserved.  The eccentricities of some of
the planets (particularly Uranus and Saturn) increase slightly, but
attain new equilibrium values.  The evolution of the eccentricities
for the system with Saturn and Uranus in the 2:3 MMR and Uranus and
Neptune in the 3:4 MMR is shown in Fig.~\ref{evap_e}.

Once the gas is removed, we can continue following the evolution of
the our systems with an N-body code, accounting only for the Sun and
the 4 planets. The simulations are done with the symplectic integrator
SYMBA (Duncan et al., 1998) and cover a timescale of 1 Gy (assuming
that $r=1$ corresponds to 5.1~AU so that the orbital period there is
11.5~y). The time-step is 0.2~y.

Note that because the hydro-dynamical simulations were carried out in
two dimensions, the planetary orbits are strictly co-planar in our
$N$-body simulations.  We do not think that this is a significant
limitation because it is well known that gas disks very
effectively damp planetary inclinations (Lubow and Ogilvie, 2001;
Tanaka and Ward, 2004) and we see no way to effectively excite them 
again.  Planetary inclinations can only be excited by close encounters
(which do not occur in our resonant configurations) or if the planets
were trapped in inclination MMRs. Inclination resonances, however, are
much weaker than the eccentricity resonances that occur at the same
location (they act as second order resonances because the inclination
has to appear with an even power in the equations of motion for
D'Alembert rules --- see Morbidelli 2002), so trapping in inclination
resonances is highly unlikely. Thus, we expect that the real planets
have small inclinations when they emerge from the gas disk, so
that the study of the long-term stability of the multi-resonant
configuration can be done effectively in two dimensions.

We find that the configuration with Saturn and Uranus in the 2:3 MMR
and Uranus and Neptune in the 3:4 MMR remains stable for the full
integration time, with no visible changes in semi-major axes or
eccentricities (Fig.~\ref{4pl1Gy}).  Remember, however, that this
simulation does not take into account the effects of a remnant
planetesimal disk, which was used in the original Nice model to drive
the planetary system unstable.  We study this situation in
section~\ref{Nice2}.
  
The configuration with Saturn and Uranus in the 2:3 MMR and Uranus and
Neptune in the 4:5 MMR remains stable for 400~My.  This time is long
enough, however, that this configuration might be a reasonable
starting point for a Nice-model-like evolution, because it is
consistent with the 650 Myr delay between planet formation and the
onset of the LHB.

All other configurations become unstable very quickly, at times
ranging from a fraction of a My (for the most compact configuration)
to 27 My (in the case with both Saturn and Uranus, and Uranus and
Neptune in the 3:4 MMR).  In these cases, however, we became concerned
that we were driving these systems unstable because our disk
dispersal time was too short.  To test this possibility, we did again the
later simulation with a gas-halving time of 1100 orbital periods at
$r=1$. The simulation was run for 10,000 orbital periods, at the end of
which the gas surface density had been reduced by a factor of
$\sim\!500$.  Even during the hydro-dynamical simulation we saw signs
of instability and the system became unstable in less than 1~My in
subsequent $N$-body simulation.

Another possibility is that the gravitational effects of a distant
planetesimal disk could stabilize planetary configurations that were
otherwise unstable.  Thus, we redid each of the aforementioned
$N$-body simulations twice: once adding a disk of 50 Earth-masses and
once adding a 80 Earth-masses disk.  In both cases, the disk was
represented by a collection of 2,000 massive particles that ranged in
heliocentric distance from just beyond the 2:3 MMR with the outermost
planet, to $\sim 30$~AU. We placed the inner edge of the disk this
far from the Sun because we did not want a significant number of
particles to leak out of the disk and trigger planetary migration.
Such a migration would drastically change the structure of the system
making a comparison with the disk-free simulations impossible.  Such a
distant disk could still irreversibly damp the planets' eccentricities
through a secular exchange of angular momentum and mixing of the
planetesimals' secular phases.  Nevertheless, in all our simulations
we found that the planetary configurations become unstable in very
short periods of time ($\sim 10$~My).

Given the above results, we believe that 4 out of the 6 relative
configurations that we found are so unstable that they could not have
lasted long enough to explain the 650 Myr delay between the formation
of the planets and the LHB.  This, however, does not preclude the idea
that other planetary systems might have passed through similar
configurations --- becoming unstable soon after the disappearance of
the gas disk.  Quite interestingly, we find that the instabilities
that characterize these systems can often be much more violent than
the one we can tolerate for the LHB, and involve close encounters
between Jupiter and Saturn.  As such, they can leave Jupiter on an
orbit at about 4--5~AU with an eccentricity comparable to that of some
extra-solar planets (see Fig.~\ref{exopl}, also see Rasio and Ford,
1996, and Mazari and Weidenschilling, 2002). This opens the
possibility that those extra-solar planets that have been found on
eccentric orbits beyond $\sim 3$~AU from their host stars might have
followed an { evolution similar to the ones that we found in our
hydro-dynamical simulations, but { that we rejected} for our solar
system, based on the LHB constraint}.

Finally, we thought it would be instructive to investigate whether it
was possible to add additional 15 Earth-mass ice giant to the system
and still produce stable configurations.  In order to have the best
chance at successfully constructing such a system, we start with the
most stable configuration produced above --- namely the one in which
Uranus is in the 2:3 MMR with Saturn and Neptune is in the 3:4 MMR
with Uranus. We place a fifth planet, of equal mass to that of the
other ice giants, between the 3:4 and 2:3 MMRs with Neptune at
$r=2.87$.  We then let the system evolve under the effects of the gas
disk. As expected, after a short period of Type~I migration towards
the Sun, the fifth planet is captured in the 3:4 MMR with Neptune. The
capture into resonance excites the eccentricities of Uranus and
Neptune, which stabilize at about 0.08 and 0.03 respectively.  These
values are about twice larger than those achieved in the simulations
with only 4 planets, reported in Fig.~\ref{U23Neptune}
and~\ref{U34Neptune}. This is due to the fact that Uranus and Neptune
have to retain the fifth planet from migrating through their mean
motion resonances. Thus they suffer an additional negative
torque. Because they are themselves locked in resonances, and
therefore cannot migrate, this translates in a stronger excitation of
their orbital eccentricities.  On the
other hand, the eccentricity of the fifth planet stays below 0.02.  

After 2,000 Jovian orbital periods, we start the procedure of
gradually depleting the gas disk using the procedure explained in the
previous section.  Unlike the previous cases, however, this system
already shows signs of being unstable during this phase of its
evolution.  In particular, we see a secular increase in the amplitude
of oscillation of both Jupiter's and Saturn's eccentricities.  In
addition, the variation in the eccentricities of all three ice giants
become noticeably erratic. All this is probably a consequence of the
enhanced eccentricities of Uranus and Neptune, relative to the
4-planets simulations. 

Once the gas is gone, we move the system to our $N$-body code.  We
find that the system becomes violently unstable on a timescale of
$\sim 10$~My. To test this result, we have performed {3}
additional $N$-body simulations starting from the output of the
hydro-dynamical code at slightly different times, separated by 100
orbital periods at $r=1$.  In all cases the planets become unstable in
less then 20~My.  We have also performed runs where we have added
either a 50 or 80 Earth-mass planetesimal disk (as described above).
Again, the results are essentially the same.  We only managed to
slightly delay the onset of the instability to $\sim 30~$My.

Of course, we cannot rule out that there was once a fifth fully grown
planet in the outer solar system based on these simulations alone.
After all, we only studied one configuration, and the fifth planet
might have had a different mass or Uranus and Neptune might have been
in different resonances then we assumed.  Furthermore, as noted in the
previous section, it is possible that other resonant configurations
may have been reached if we changed the structure of the gas disk.
Finally, there is a chance that the system was more unstable than it
might have been because we dispersed the gas disk too quickly.
Nevertheless, the striking difference between the behavior of our
five-planet system and the four-planet case that has Uranus and
Neptune in the same MMRs suggests that it is probably much more
difficult to find a five-planet configuration that is stable for a
long period of time.

\section{From a fully resonant evolution to the current orbital
  architecture}
\label{Nice2}

We conclude this paper by presenting a couple of `proof-of-concept'
N-body simulations in which the four giant planets interact with a
trans-planetary disk of planetesimals that is close enough to the
planets to cause them to migrate.  The results demonstrate that the
multi-resonant planetary systems described above can indeed evolve
into an orbital configuration similar to that of the real giant
planets.

We start with the most stable of our multi-resonant systems, namely
the one in which Jupiter and Saturn are in the 2:3 MMR, Saturn and
Uranus are in the 2:3 MMR, and Uranus and Neptune in the 3:4 MMR.  As
we have seen in Fig.~\ref{4pl1Gy}, this system is stable for at least
a billion years in the absence of external perturbations.  We now add
a trans-planetary disk of planetesimals.  As in Tsiganis et al.\
(2005), we place the inner edge of the disk close to the outermost
planet (0.5~AU beyond it), so that the planets migrate very quickly.
This was a purely practical decision that allowed us to save a
significant amount of computing time.  As in Tsiganis et al. (2005),
we model the disk with 1,000 equal-mass planetesimals, with a surface
density profile that is inversely proportional to heliocentric
distance, $\Sigma (r)\sim 1/r$.  The outer edge of the disk is placed
30~AU, and its total mass is set to 50 Earth masses.  All particles
are initially on nearly circular and co-planar orbits with
$e\sim\sin{(i)}\sim 10^{-3}$.

The initial conditions of the four planets are based on the output of
the hydro-dynamical simulation with a decreasing gas disk. However, since our
hydro-dynamical simulations were performed in two dimensions, they
output only a co-planar configuration.  This limitation is acceptable
as long as the orbits of the planets do not cross one another.
However, if they do cross (as we expect in these simulations) the 2D
assumption artificially increases the chances of a collision to an
unacceptable level.  To avoid this technical problem, we add a small
($\sim 10^{-3}$~AU/y) $z$-component to the velocity vector of each planet
at the beginning of these $N$-body simulations.  We followed the
evolution of each system for 100~My using SyMBA.

The result for our first run, shown in Fig.~\ref{rescros1}, has many
of the characteristics seen in Fig.\ (1) of Tsiganis et al.\ (2005).
In particular, the planets undergo a short period of smooth migration,
during which they are on circular orbits.  This is followed by an
abrupt increase in the eccentricities of the gas giants, which
destabilizes the orbits of the ice giants, and leads to a short, but
violent, period of repeated encounters between the planets. Then there
is a period when the planets migrate very quickly through the
remaining disk, while their eccentricities slowly decay due to
dynamical friction. The planets reach their final orbits in $\sim
100~$My, when the planetesimals disk has been dispersed.

The essential ingredients of the Nice model are preserved in these new
simulations. The initial orbits of the planets are stable and thus the
instability does not occur until the planets are forced to migrate
across a MMR.  In the original Nice model the instability was caused
by Jupiter and Saturn crossing the 1:2 MMR. Here, since Saturn starts
off much closer to the Sun, the first important resonance that the
planets encounter is the 3:5 MMR between Jupiter and Saturn.  As
Fig.~\ref{rescros1} shows, this crossing causes the instability.

Using a perturbation theory similar to the one in the supplementary
material of Tsiganis et al.~(2005), one can show that the 3:5 MMR is
less effective than the 1:2 MMR at increasing the eccentricities of
the gas giants.  However, since the planetary configuration that we
used here is more compact than that in the original Nice model, a mild
eccentricity ``jump'' is enough to destabilize the orbits of the ice
giants.  Thus, as we suggested in section~\ref{4pl-gas}, we can
conclude that in order to reproduce a Nice-model-like instability in
the orbits of the ice giants of a system that initially has Jupiter
and Saturn locked in the 2:3 MMR, Uranus and Neptune probably need to
be closer to the Sun than the original Nice model postulated.

In the run shown in Fig.~\ref{rescros1}, Uranus and Neptune suffer a
few encounters with each other before the latter is scattered into the
disk (at about 15~AU).  Dynamical friction from the disk decouples
Neptune from Uranus, after which it migrates smoothly outward on a
nearly circular orbit.  Neptune stops migrating when it hits the outer
edge of the disk. Notice that Uranus does not migrate far enough since
its final semi-major axis is $\sim 17$~AU instead of 19.2~AU.  This
behavior is reminiscent of the subset of simulations from Tsiganis et
al.\ (2005), where the ice giants do not encounter Saturn.  Indeed,
the location of Uranus was one reason why Tsiganis et al.\ concluded
that such encounters must have happened.

However, at the instability time, a number of different behaviors are
possible due to the chaotic nature of the dynamics. For example, 
in a second simulation that is similar to the one in
Fig.~\ref{rescros1}, but with a 65 Earth-mass disk,
(Fig.~\ref{rescros2}) Saturn is involved in
gravitational encounters with the ice giants.  As a result, Neptune is
thrown much farther into the disk, landing at $a\sim 25$~AU on a very
eccentric orbit.  Its orbit is subsequently circularized by dynamical
friction, and it comes to rest in a nearly circular orbit near 30~AU.
In addition, Uranus's final semi-major axis is very close to its
observed value.  This result is also consistent with the findings of
Tsiganis et al.\ (2005).  Notice that, in this run, the ice giant that
formed closer to the Sun became the most distant planet in the final
system.

As described above, in the original Nice model the orbital instability
was cause by Jupiter and Saturn crossing the 1:2 MMR, while in these
simulations it is caused by the 3:5 MMR.  The 3:5 MMR is closer to the
Sun then the 1:2 MMR, and, since Saturn is currently found beyond
Jupiter's 1:2 MMR, it eventually had to cross it.  This
resonance-crossing excites, once more, the gas giants' eccentricities,
and thus helps them maintain non-zero values against dynamical
friction.  We note that by the time Jupiter and Saturn cross their 1:2
MMR, the mass of the remaining disk in our runs is roughly 25
Earth-masses -- a value that is well within the range needed to
explain the capture of Jupiter Trojans during the 1:2 MMR crossing,
according to the model in Morbidelli et al$.$~(2005).

Of course, much more work is needed to build a successful `Nice
model~II' that starts from an initial multi-resonant configuration of
the giant planets.  It is crucially important, for example, to
determine whether it is possible to delay the instability for 650 Myr
in order to be consistent with the LHB chronology.  Recall that in the
above simulations, we purposely set the disk's initial distribution so
that the resonance crossing occurs early in order to save CPU time.
However, as discussed in Gomes et al. (2005), a more realistic
distribution of the planetesimal disk should contain only particles
whose dynamical lifetime is of order of the lifetime of the gas disk
(few My) or longer. Assuming this disk distribution, Gomes et
al. showed that, at least for the initial planetary configuration
assumed in the original Nice model, the migration of the planets is
slow enough that the instability is achieved only after hundreds of
millions of years, consistent with the LHB timing. The same would
hopefully happen for our new initial planetary configuration.
Moreover, a large number of simulations need to be performed in order
to quantify the probability that the final orbits achieved by the
planets from our new initial configuration are consistent with
observations.  This study, which will require many time-consuming
simulations is currently ongoing, and will be the subject of a
forthcoming paper.

\section{Concluding remarks}
\label{conclusions}

The are two important characteristics of our solar system that any
model must be able to explain.  First, the hot and warm Jupiters that
are seen around some other stars are not present in our system.
Second, the Moon and the planets most likely carry the scars of a
spike in the impact flux that occurred $\sim\!650$~My (the Late Heavy
Bombardment or LHB) after planets formed. The LHB strongly suggests
that the planets suddenly became unstable at that time, destabilizing
a massive reservoir of small bodies (Levison et al., 2001).

The so-called Nice model (Tsiganis et al., 2005; Gomes et al., 2005)
has been proposed, in part, to explain the LHB.  This model has a long
list of successes in reproducing many of observational characteristics
of the solar system. These include the number and the orbital
distribution of the Jovian Trojans (Morbidelli et al., 2005) and the
irregular satellites of Saturn, Uranus and Neptune (Nesvorn\'y et al.,
2007), and the structure of the Kuiper belt (Levison et al., 2007).
The main weakness of the Nice model is that the initial conditions of
the planets were chosen without concern for the previous phase of
planetary evolution when the the proto-planetary gas disk was still in
existence.

Recent hydro-dynamic simulations of Jupiter and Saturn embedded in a
gas disk have supplied an important clue about the initial stable
configuration of the planets.  In particular, MS01 and MC07 showed
that the absence of a hot/warm Jupiter in our system can be explained
if these two planets had been locked in their mutual 2:3 MMR.  The
main goal of this paper, therefore, is to extended these results and
find a four planet configuration that is both non-migrating while the
gas disk is present, and dynamically stable long after the gas disk
disperses.

To accomplish this goal, we have taken a system from MC07 consisting
of Jupiter, Saturn, and a gaseous disk, and progressively added Uranus
and Neptune, one at a time, into the simulations.  We find that the
interaction with the gas disk drives these planets into a
configuration where each is in a MMR with its immediate neighbor(s).
We have found six such ``fully resonant'' configurations, all of which
are characterized by, at most, a small amount of radial migration.
Four configurations, however, become rapidly unstable (on a timescale
of a few My) after the disappearance of the gas disk. The remaining
two were the least compact systems, with Saturn and Uranus in the 2:3
MMR and Uranus and Neptune in either the 4:5 or the 3:4 MMR.  They
were stable for 400 Myr and over a Gyr, respectively.  

Furthermore, we have presented a pair `proof-of-concept' simulations,
showing that a quadruple resonant configuration, like the stable one
above, can evolve into a system with a structure similar to that
observed in the real solar system, once it interacts with a suitable
trans-planetary planetesimal disk.  The system evolves as follows.
The migration of the giant planets, induced by the interaction with
this disk, increases the ratios of the orbital periods between each
pair of planets. Thus, the planets are extracted from their mutual,
quadruple resonance. Because the system is very compact, new
resonances are crossed during the migration.  These resonance excite
the eccentricities of the planets, triggering a global instability of
the system. The orbits of the planets are eventually stabilized by the
dynamical friction exerted by the planetesimal disk during its
dispersal.  Thus, this evolution is different from the original Nice
model (Tsiganis et al.~2005) in the technical details only (e.g.
different resonances are involved), but not in terms of the basic
dynamical processes at work. More work is needed in order to quantify
the statistical outcome of the chaotic evolution of the planets and,
particularly, to prove that the onset of the planetary instability can
occur late, as in Gomes et al.\ (2005). This will be the object of a
forthcoming paper.

The long-term aim of this research is to build a bridge between our
knowledge on solar system dynamics during the gas-disk era and that
during the planetesimal-disk era, which remained up to now totally
disconnected to each other.  Success in this task would represent a
significant advancement for our understanding of planet formation and
of the key processes that made our solar system so different from all
extra-solar systems discovered so far.  In this paper we have taken
the first steps toward this goal.

\acknowledgments
{\bf Acknowlesgments:} A.M. and K.T. are grateful for the support received through the
France-Greece scientific collaboration programme. A.M. also
acknowledges support from the French National Programme of Planetary
science (PNP). Computations for this paper have been done on the 'Mesocentre
SIGAMM' machine, hosted by Observatoire de la Cote d'Azur. 
We also thank reviewer D. Richardson for constructive comments.

\centerline\textbf{ REFERENCES}

\begin{itemize}

\item[] Adams, F.~C., Hollenbach, 
D., Laughlin, G., Gorti, U.\ 2004.\ Photoevaporation of Circumstellar Disks 
Due to External Far-Ultraviolet Radiation in Stellar Aggregates.\ 
Astrophysical Journal 611, 360-379. 

\item[] Alexander, R.~D., 
Clarke, C.~J., Pringle, J.~E.\ 2006.\ Photoevaporation of protoplanetary 
discs - II. Evolutionary models and observable properties.\ Monthly Notices 
of the Royal Astronomical Society 369, 229-239. 

\item[] Crida, A., Morbidelli, A., Masset, F. 2007. Simulating planet migration in globally evolving 
disks.\ Astronomy and Astrophysics 461, 1173-1183.

\item[] Duncan, M.~J., Levison, 
H.~F., Lee, M.~H.\ 1998.\ A Multiple Time Step Symplectic Algorithm for 
Integrating Close Encounters.\ Astronomical Journal 116, 2067-2077. 

\item[] Gomes, R.~S., Morbidelli, 
A., Levison, H.~F.\ 2004.\ Planetary migration in a planetesimal disk: why 
did Neptune stop at 30 AU?.\ Icarus 170, 492-507. 

\item[] Gomes, R., Levison, 
H.~F., Tsiganis, K., Morbidelli, A.\ 2005.\ Origin of the cataclysmic Late 
Heavy Bombardment period of the terrestrial planets.\ Nature 435, 466-469. 

\item[] Guillot, T., Hueso, 
R.\ 2006.\ The composition of Jupiter: sign of a (relatively) late 
formation in a chemically evolved protosolar disc.\ Monthly Notices of the 
Royal Astronomical Society 367, L47-L51. 

\item[] Haisch, K.E., Lada, E.A. and Lada, C.J. 2001. Disk frequencies
  and lifetimes in young clusters. Astron. J., 553, L153-156. 

\item[] Hayashi C. 1981. Structure of the solar nebula, growth and decay of magnetic
fields and effects of magnetic and turbulent viscosities on the
nebula. {\it Prog. Theor. Phys. Suppl.}, {\bf 70}, 35-53.

\item[] Levison, H.~F., Dones, 
L., Chapman, C.~R., Stern, S.~A., Duncan, M.~J., Zahnle, K.\ 2001.\ Could 
the Lunar ``Late Heavy Bombardment'' Have Been Triggered by the Formation 
of Uranus and Neptune?.\ Icarus 151, 286-306. 

\item[] Levison, H.F., Morbidelli A., gomes, R. and Tsiganis K. 2007. 
Origin of the structure of the Kuiper belt during a dynamical
instability in the orbits of Uranus and Neptune. Icarus, submitted.

\item[] Lubow, S.~H., 
Ogilvie, G.~I.\ 2001.\ Secular Interactions between Inclined Planets and a 
Gaseous Disk.\ Astrophysical Journal 560, 997-1009. 

\item[] Marzari, 
F., Weidenschilling, S.~J.\ 2002.\ Eccentric Extrasolar Planets: The 
Jumping Jupiter Model.\ Icarus 156, 570-579. 

\item[] Masset, F.\ 2000a.\ FARGO: A 
fast eulerian transport algorithm for differentially rotating disks.\ 
Astronomy and Astrophysics Supplement Series 141, 165-173. 

\item[] Masset, F.~S.\ 2000b.\ FARGO: A 
Fast Eulerian Transport Algorithm for Differentially Rotating Disks.\ ASP 
Conf.~Ser.~219: Disks, Planetesimals, and Planets 219, 75. 

\item[] Masset, F., 
Snellgrove, M.\ 2001.\ Reversing type II migration: resonance trapping of a 
lighter giant protoplanet.\ Monthly Notices of the Royal Astronomical 
Society 320, L55-L59. 

\item[] Morbidelli, A., 
Levison, H.~F., Tsiganis, K., Gomes, R.\ 2005.\ Chaotic capture of 
Jupiter's Trojan asteroids in the early Solar System.\ Nature 435, 462-465.

\item[] Morbidelli A., 2002. {\it Modern Celestial Mechanics:
aspects of Solar System dynamics}, in ``Advances in Astronomy and
Astrophysics'', Taylor \& Francis, London.

\item[] Morbidelli, A. and Crida, A. 2007. The dynamics of Jupiter and
  Saturn in the gaseous proto-planetary disk. Icarus, in press.

\item[] Morbideli A., Levison, H.F. and Gomes, R. 2007. The dynamical
  structure of the Kuiper belt and its primordial origin. in {\it
  The Solar System beyond Neptune}, A. Barucci et al. eds., University
  of Arizona press, in press. 

\item[] Nesvorn{\'y}, D., 
Vokrouhlick{\'y}, D., Morbidelli, A.\ 2007.\ Capture of Irregular 
Satellites during Planetary Encounters.\ Astronomical Journal 133, 
1962-1976. 

\item[] Petit, J.-M., Chambers, 
J., Franklin, F., Nagasawa, M.\ 2002.\ Primordial Excitation and Depletion 
of the Main Belt.\ Asteroids III 711-723. 

\item[] Rasio, F.~A., Ford, 
E.~B.\ 1996.\ Dynamical instabilities and the formation of extrasolar 
planetary systems.\ Science 274, 954-956. 

\item[] Shakura, N.~I., 
Sunyaev, R.~A.\ 1973.\ Black holes in binary systems. Observational 
appearance..\ Astronomy and Astrophysics 24, 337-355.

\item[] Tanaka, H., Ward, 
W.~R.\ 2004.\ Three-dimensional Interaction between a Planet and an 
Isothermal Gaseous Disk. II. Eccentricity Waves and Bending Waves.\ 
Astrophysical Journal 602, 388-395. 

\item[] Tsiganis, K., Gomes, 
R., Morbidelli, A., Levison, H.~F.\ 2005.\ Origin of the orbital 
architecture of the giant planets of the Solar System.\ Nature 435, 
459-461. 

\end{itemize}

\newpage
\centerline{Figure captions}

\begin{itemize}
\item[Fig.~\ref{MC07fig}] The evolution of Jupiter and Saturn in the gas disk
  ($H/r=0.05$, $\nu=3.2\times 10^{-6}$) after they became locked in their
  mutual 2:3 MMR. From MC07.
\item[Fig.~\ref{Uranus}] The evolution of Uranus, after it was
  added to the simulation presented in Fig.~\ref{MC07fig}. Four
  simulations are presented. In the first (black line, labelled
  `Uranus sim.\#1'), Uranus is started beyond the 1:2 MMR with
  Saturn. During its inward migration it passes across the 1:2 MMR and
  eventually gets trapped into the 2:3 MMR. In the second simulation
  (dark gray curve, labelled sim.\#2), Uranus is introduced between
  the 2:3 and 1:2 MMRs with Saturn. Again, Uranus 
  and gets captured into the 2:3 MMR. In the third simulation 
  (gray line, labelled
  `Uranus sim.\#3'), Uranus is started between the 3:4 and the 2:3
  MMRs with Saturn and gets captured in the former. In the fourth
  simulation (light gray line, labelled `Uranus sim.\#4') Uranus is
  started between the 4:5 and 3:4 MMRs and it evolves outward until it
  is captured again in the 3:4 MMR. The evolution of Jupiter and
  Saturn is essentially the same in all the  simulations, so we only
  present one example. 
\item[Fig.~\ref{U23Neptune}] The evolution of Neptune, after it was
  added to the simulation in which Uranus and Saturn are in the 2:3
  MMR.  Four simulations are presented. In the first one (black line,
  labelled `Neptune sim.\#1') Neptune is started beyond the Uranus 2:3
  MMR, but jumps over it and gets captured in the Uranus 3:4 MMR. In the
  second one (dark gray line, labelled `Neptune sim.\#2'), Neptune is
  started between the 3:4 and the 2:3 MMRs with Uranus and gets
  captured in the former. In the third simulation (gray line, labelled
  `Neptune sim.\#3'), Neptune is started between the 4:5 and 3:4 MMRs
  and is captured in the former. In the fourth simulation (light gray
  line, labelled `Neptune sim.\#4'), Neptune is started between the 5:6
  and 4:5 MMRs and is captured in the former. The evolution of
  Jupiter, Saturn and Uranus is essentially the same in the four
  simulations, so we only
  present one example.
\item[Fig.~\ref{U34Neptune}] The same as Fig.~\ref{U23Neptune}, but
  for the case where Neptune is added to the simulation in which
  Uranus is in the 3:4 MMR with Saturn.
\item[Fig.~\ref{evap_e}] The evolution of the eccentricities of Uranus
  (top, gray curve), Saturn (second from the top, dark gray curve),
  Neptune (third from the top, light gray curve) and Jupiter (bottom,
  black curve), during the simulation in which Uranus is in the 2:3
  MMR with Saturn and Neptune is in the 3:4 MMR with Uranus. The
  surface density of the disk is now halved every 160 orbital periods at
  $r=1$.
\item[Fig.~\ref{4pl1Gy}] The evolution of the giant planets over 1Gy,
  according to a N-body simulation, starting from the final output of
  Fig.~\ref{evap_e}. The top panel shows the evolution of the
  eccentricities with the same color code of Fig.~\ref{evap_e}. The
  bottom panel shows the evolution of the semi-major axes.   The
  evolution looks perfectly regular and stable. No planetesimal disk
  is considered, and hence no migration of the planets relative to
  each other is observed.
\item[Fig.~\ref{exopl}] The semi-major axis vs. eccentricity
  distribution of the extra-solar planets discovered by radial
  velocity technique. The size of each dot is proportional to the
  planet's radius (simply estimated from the cubic root of its $M
  sin(i)$). The rhombus shows the final orbit of Jupiter, achieved in the
  N--body simulation starting from the configuration with Uranus and
  Neptune both in the 3:4 MMR with the immediately interior
  planet. The excitation of Jupiter's orbit is due to a strong
  encounter which ejects Saturn onto a very elongated orbit. 
\item[Fig.~\ref{rescros1}] (top) Evolution of $a$, $q$, and $Q$ for
  the outer planets, under the effects of a 50 Earth-masses
  planetesimals disk.  {For illustrative purposes, the unit of
  distance used in the hydro-dynamical simulations, is scaled such
  that Jupiter has initially a semi-major axis of $5.42~$AU}. The unit
  of time is the year. The
  resonance crossing events are marked by vertical lines. The 3:5 MMR
  between Jupiter and Saturn is crossed at $t\approx 6.5~$My, while
  the 1:2 MMR is crossed much later (at $t\approx 63~$My). In this run
  Uranus falls $\sim 2~$AU short of its true location. (middle)
  Eccentricity evolution for Jupiter and Saturn. The two excitation
  episodes are clearly seen, as is the slow damping due to dynamical
  friction. (bottom) A closer look at the instability-onset phase.
  The evolution of the semi-major axes of Saturn, Uranus and Neptune
  are shown in the interval $1-30~$My. Also the evolution of the
  locations of the 3:5 MMR between Jupiter and Saturn and the 2:3 MMR
  between Uranus and Neptune are shown (time in log scale). Small
  variations in Uranus' $e$ are visible, produced by the crossing of
  high-order resonances with Neptune. However, not even their mutual
  2:3 crossing destabilizes their orbits.  This occurs exactly after
  the crossing of the 3:5 MMR between Jupiter and Saturn, at $t\approx
  6.5~$My.  After this event, Uranus and Neptune have repeated
  encounters and planetary migration is accelerated.
\item[Fig.~\ref{rescros2}] The same as Fig.~\ref{rescros1}, for the
  run with a disk of 65 Earth masses (time in log scale). The crossing
  of the 3:5 MMR between Jupiter and Saturn occurs at $t\approx
  2~$My. Then, the innermost ice giant suffers close encounters with
  both Saturn and Jupiter, which receive a ``kick'' in $a$ that forces
  them to cross their mutual 1:2 MMR. The eccentricity of the ice
  giant grows to $\sim 0.6$. Repeated encounters between the two ice
  giants follow, resulting to exchange the heliocentric ordering of
  their orbits. The planets are stabilized at 5.2, 9.2, 20 and
  32~AU. The final eccentricities of Jupiter and Saturn are $0.03$ and
  $0.07$, respectively.
\end{itemize} 

\newpage 
\begin{figure}[t!]
\begin{center}
\includegraphics[height=10.cm]{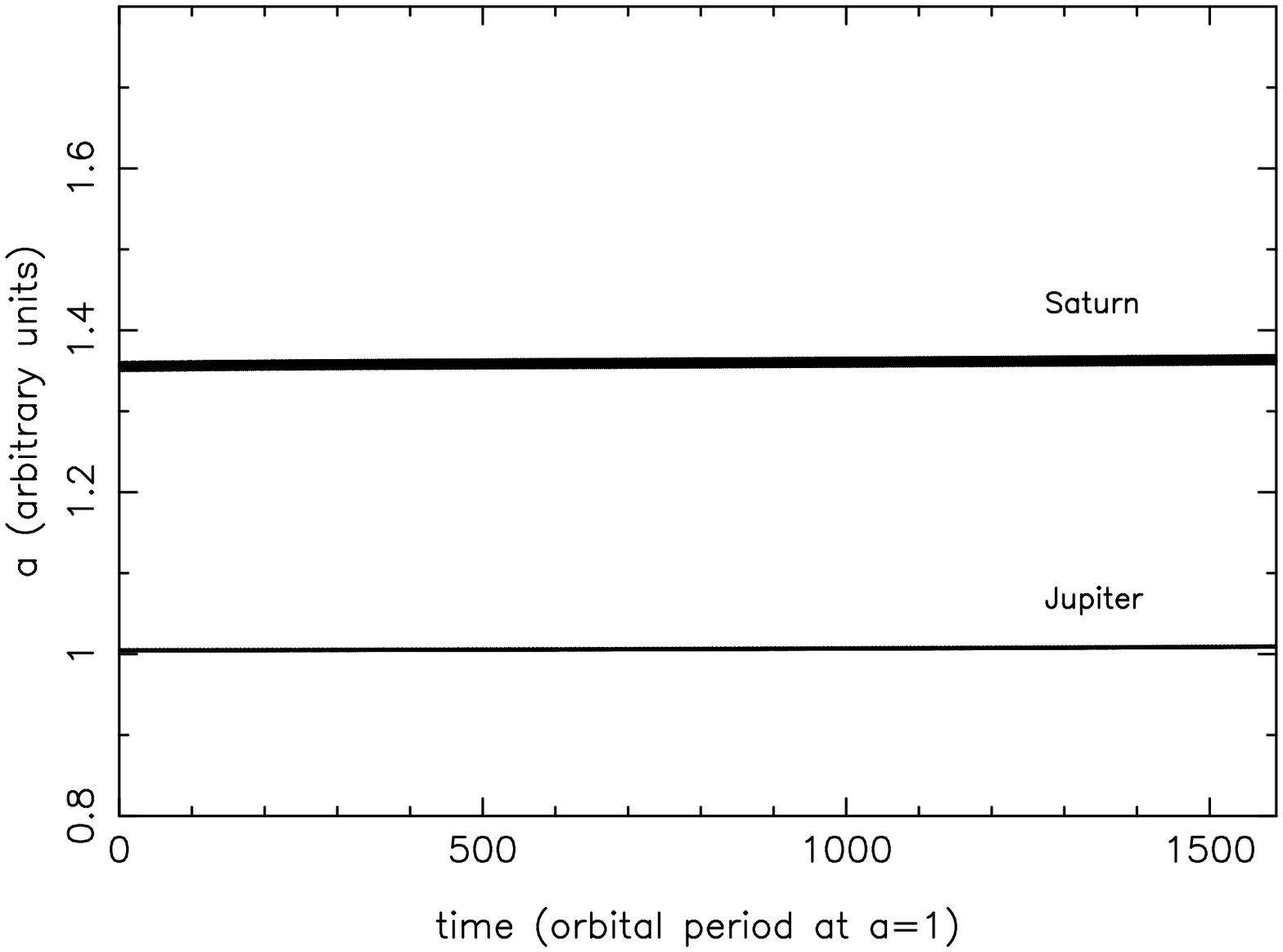}
\end{center}
\vspace*{-.3cm} 
\caption{} 
\label{MC07fig}
\end{figure} 

\newpage 
\begin{figure}[t!]
\begin{center}
\includegraphics[height=10.cm]{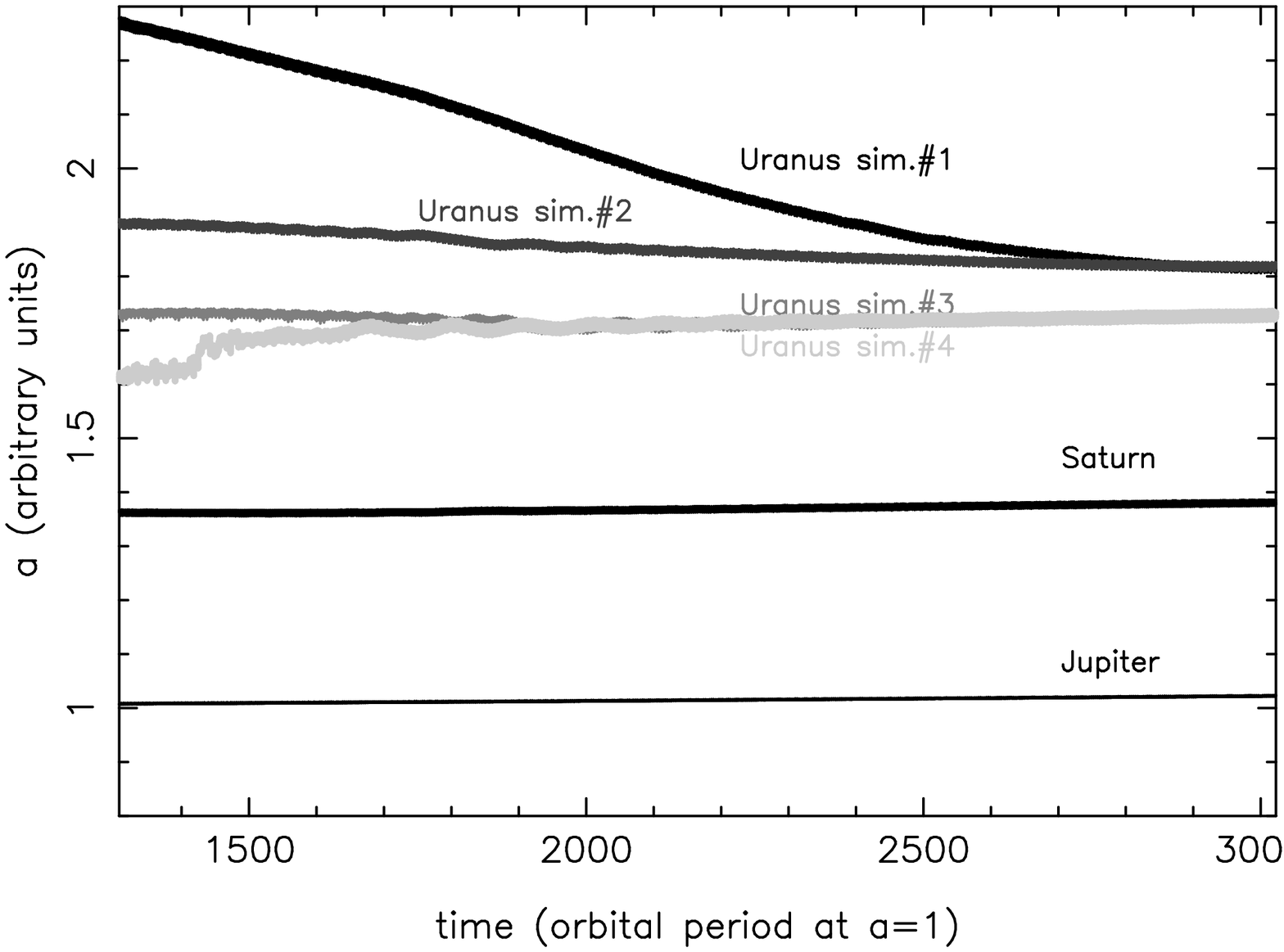}
\end{center}
\vspace*{-.3cm} 
\caption{} 
\label{Uranus}
\end{figure} 

\newpage 
\begin{figure}[t!]
\begin{center}
\includegraphics[height=10.cm]{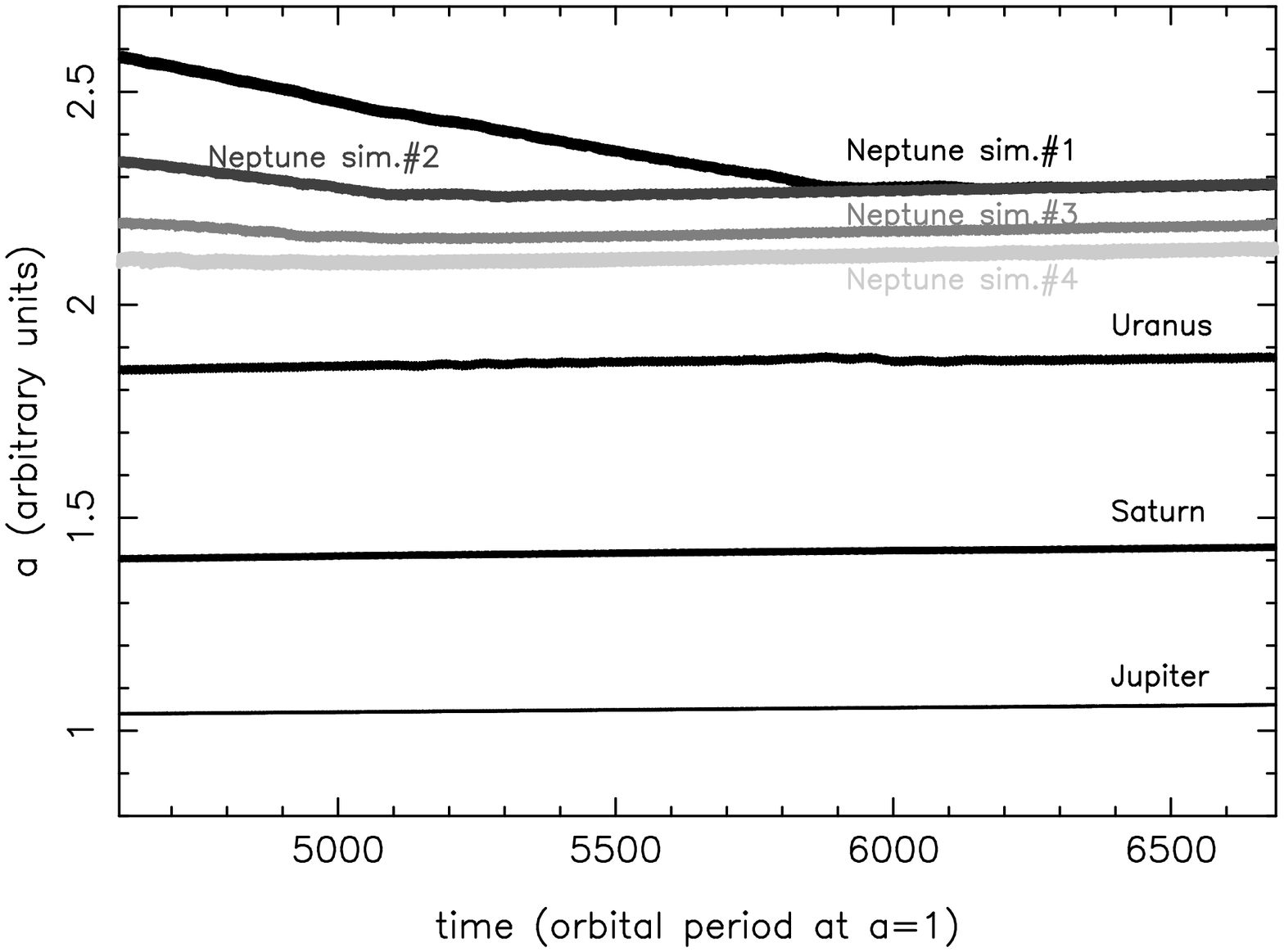}
\end{center}
\vspace*{-.3cm} 
\caption{}
\label{U23Neptune} 
\end{figure} 

\newpage
\begin{figure}[t!]
\begin{center}
\includegraphics[height=10.cm]{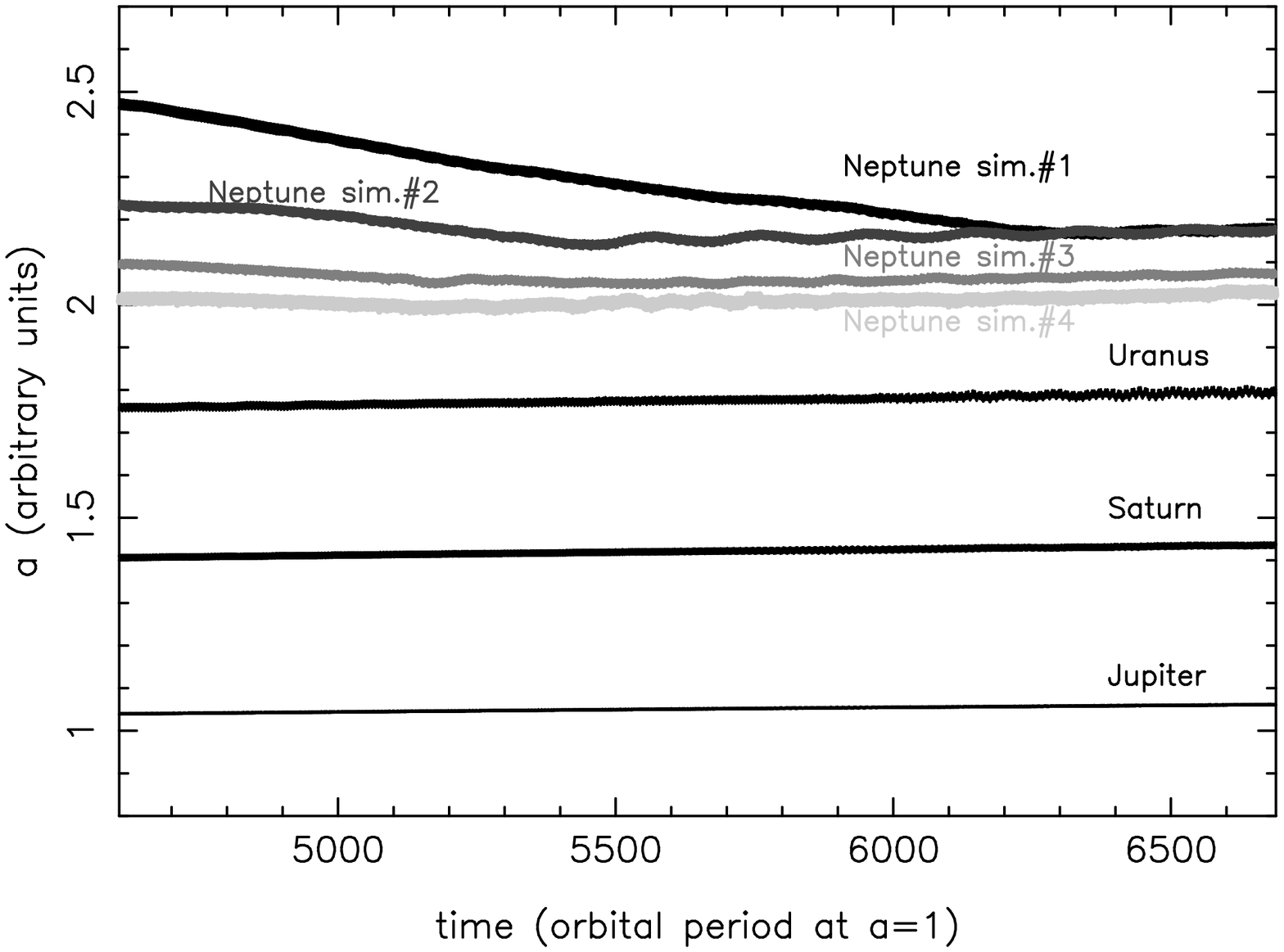}
\end{center}
\vspace*{-.3cm} 
\caption{}
\label{U34Neptune} 
\end{figure} 

\newpage
\begin{figure}[t!]
\begin{center}
\includegraphics[height=10.cm]{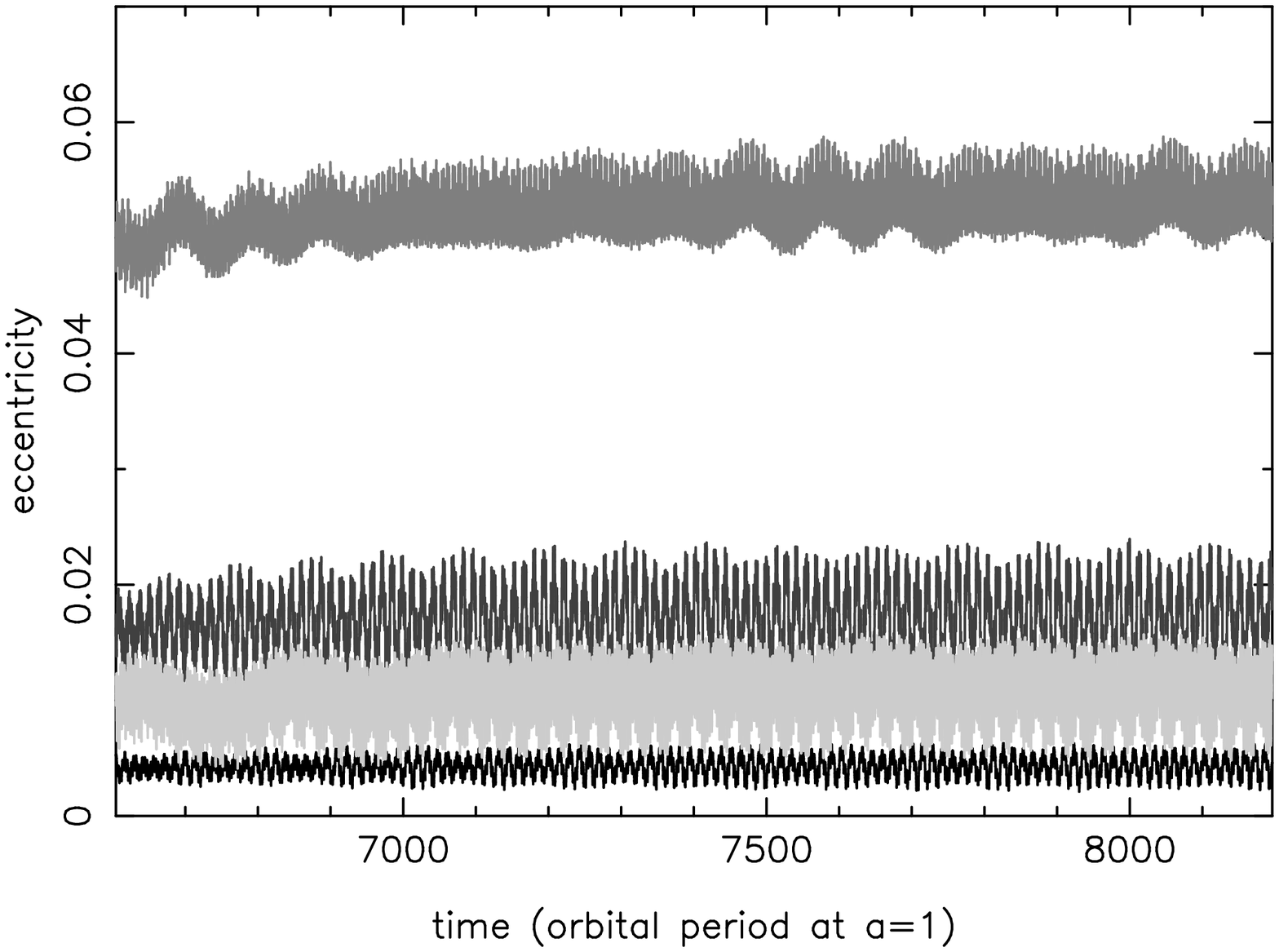}
\end{center}
\vspace*{-.3cm} 
\caption{} 
\label{evap_e}
\end{figure} 

\newpage 
\begin{figure}[t!]
\begin{center}
\includegraphics[height=10.cm]{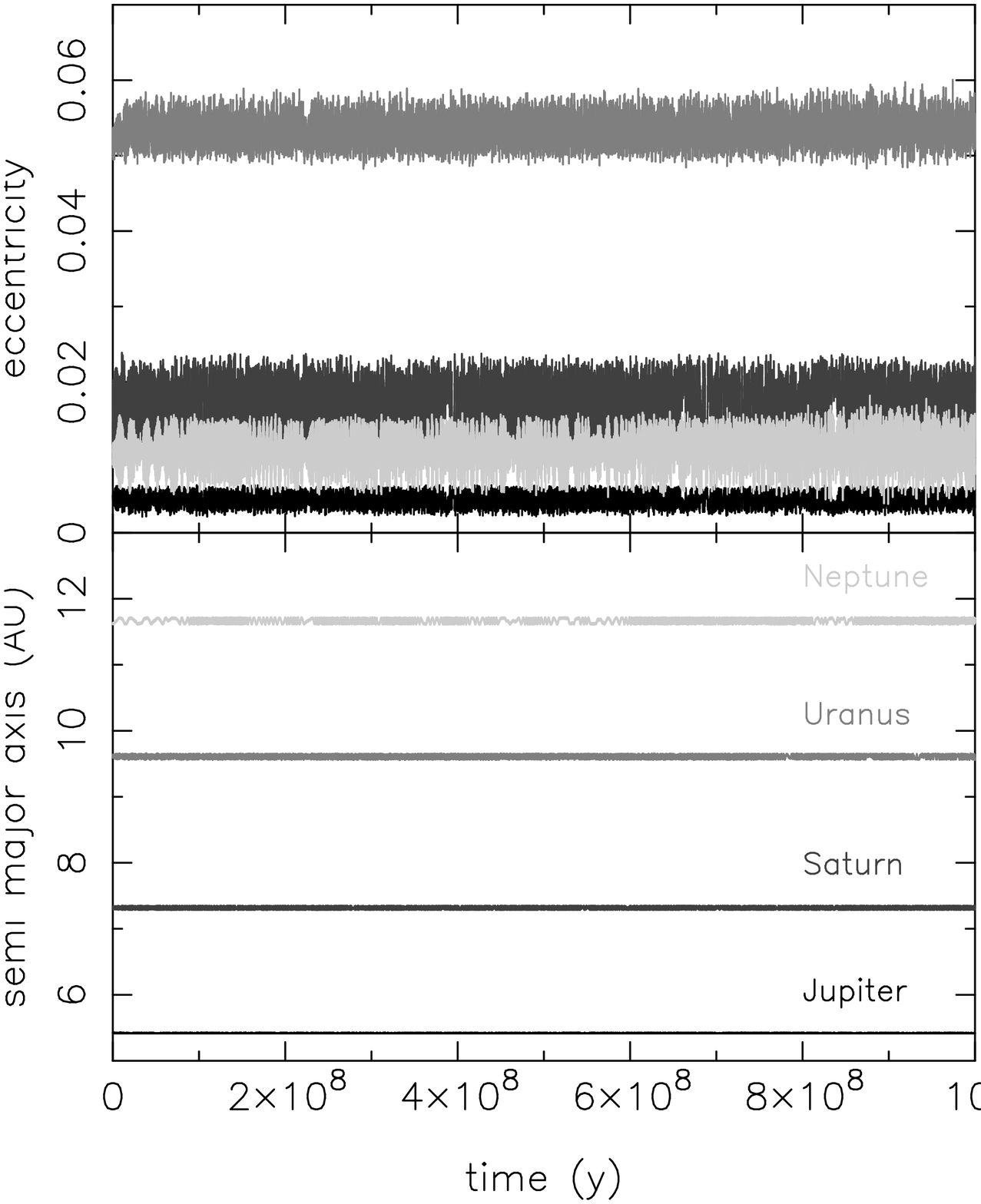}
\end{center}
\vspace*{-.3cm} 
\caption{} 
\label{4pl1Gy}
\end{figure} 

\newpage 
\begin{figure}[t!]
\begin{center}
\includegraphics[height=10.cm]{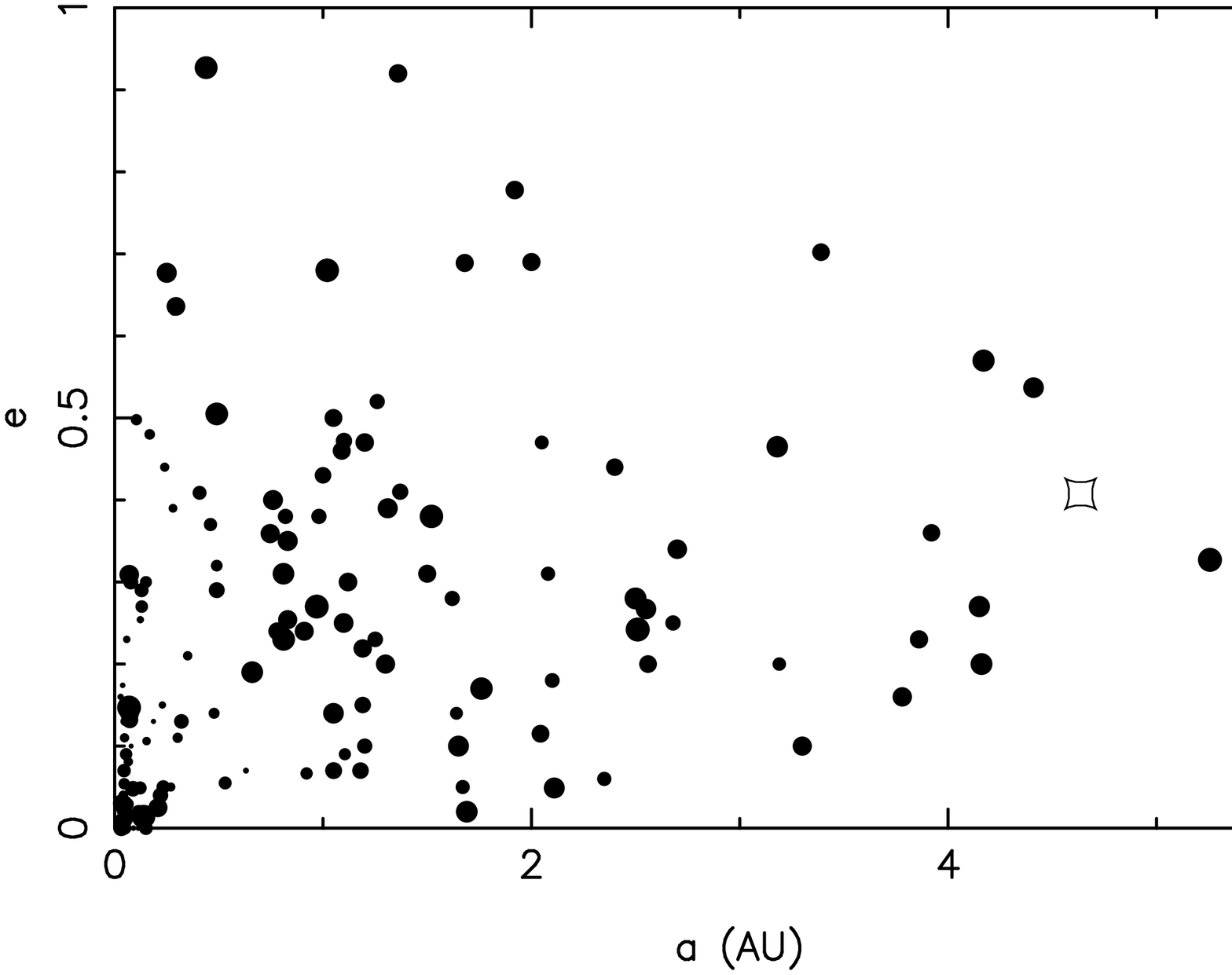}
\end{center}
\vspace*{-.3cm} 
\caption{} 
\label{exopl}
\end{figure}

\newpage 
\begin{figure}[t]
\begin{center}
\includegraphics[angle=-90,width=10.cm]{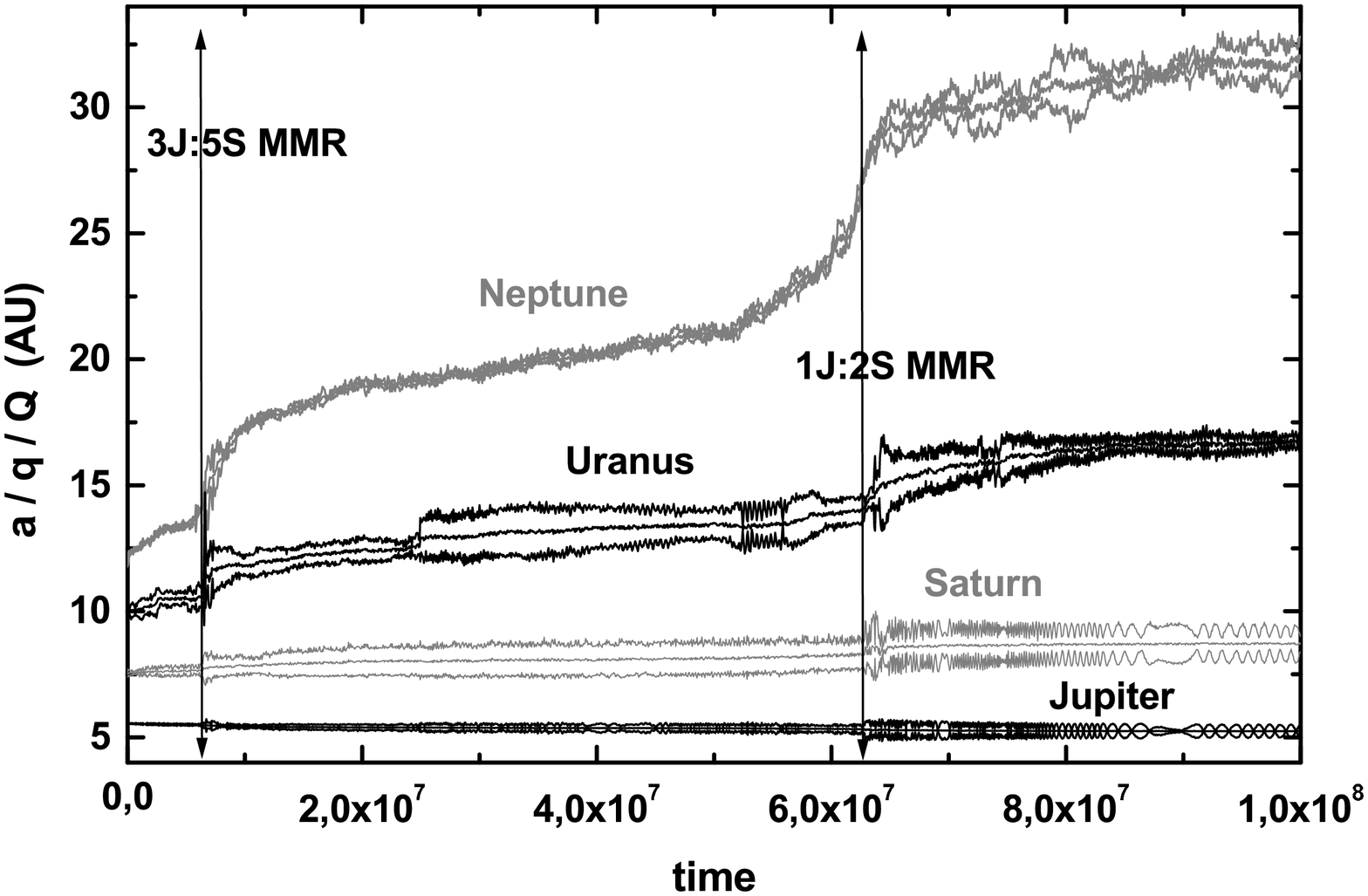}
\end{center}
\begin{center}
\includegraphics[angle=-90,width=10.cm]{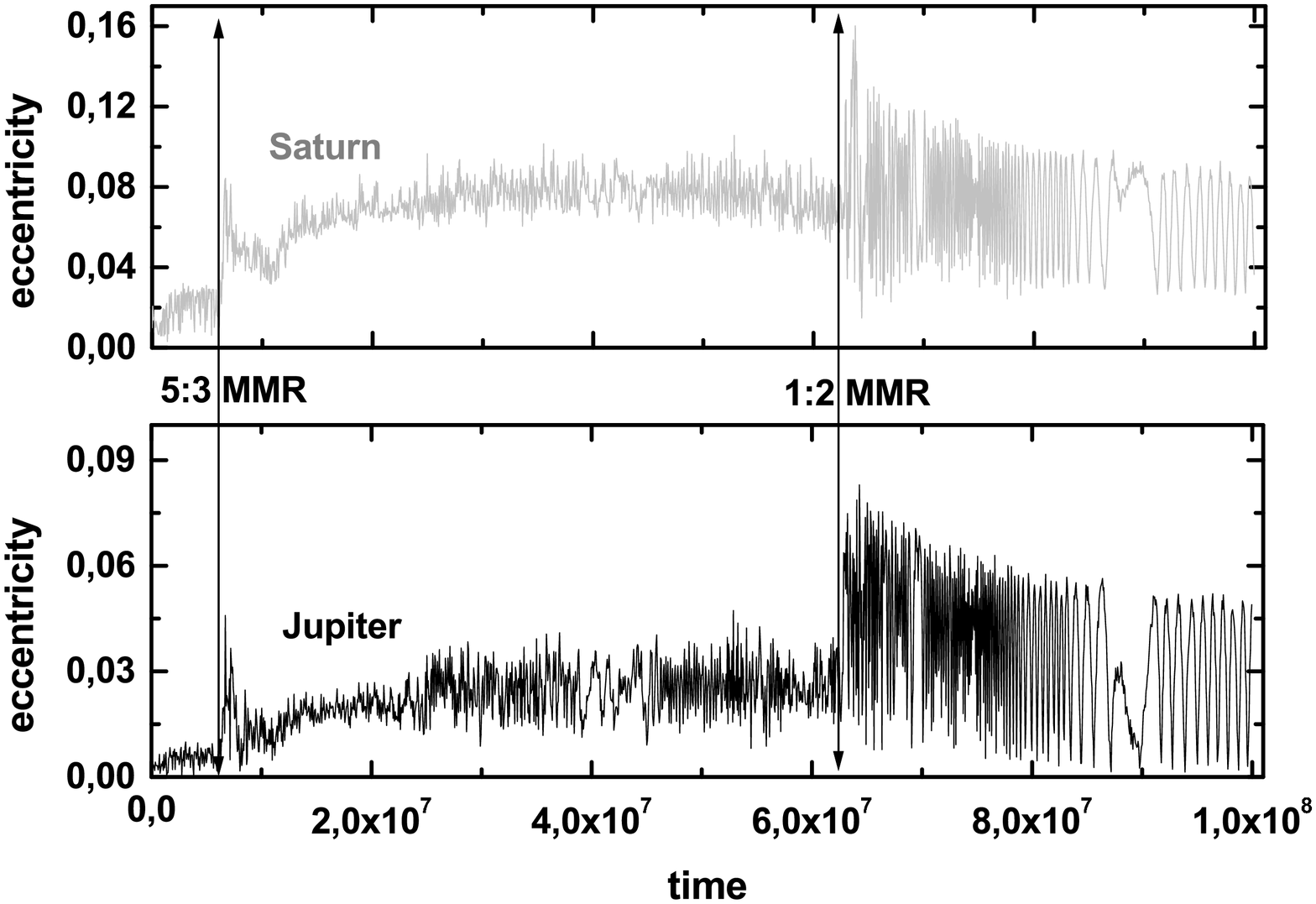}
\end{center}
\begin{center}
\includegraphics[angle=-90,width=10.cm]{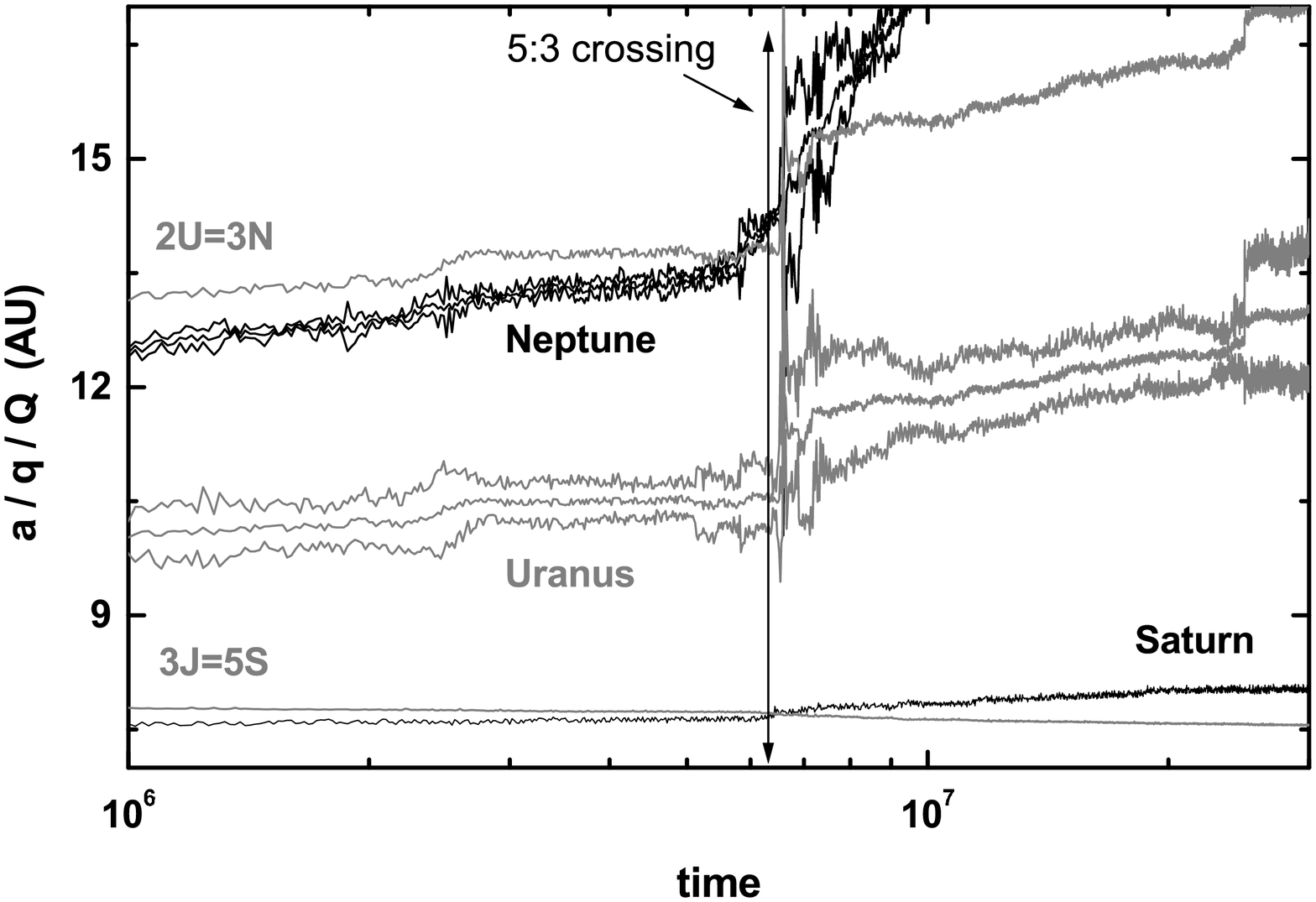}
\end{center}
\caption{} 
\label{rescros1}
\end{figure} 

\newpage 
\begin{figure}[t]
\begin{center}
\includegraphics[angle=-90,width=12.cm]{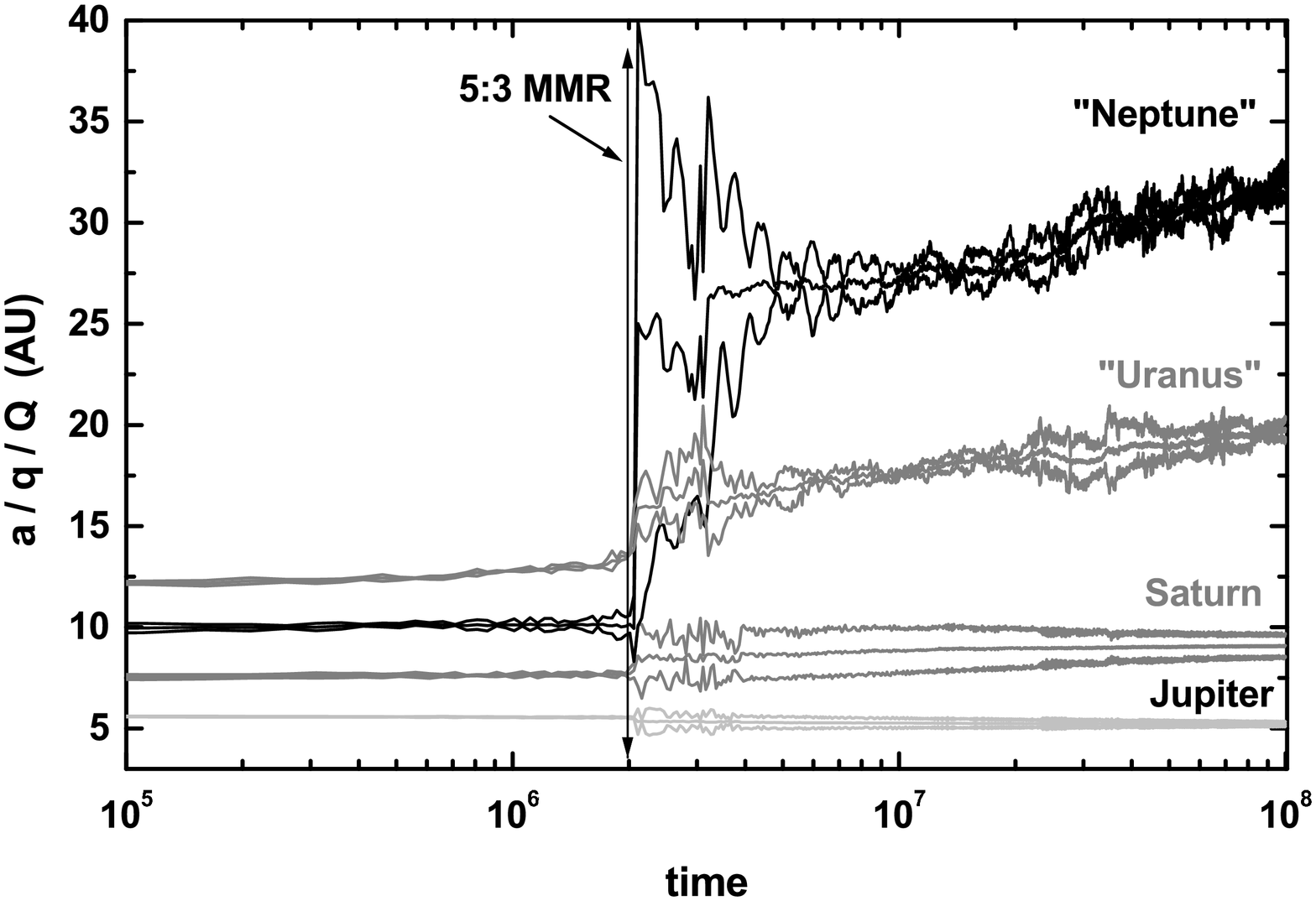}
\end{center}
\caption{} 
\label{rescros2}
\end{figure}

\end{document}